\documentstyle[twoside,epsfig,fleqn,espcrc2]{article}
\title{\bf Bose-Einstein to BCS Crossover Picture for
High-$T_{c}$ Cuprates}
\author{Y.~J.~Uemura\address{Dept. of Physics, 
Columbia University, New York, NY 10027, U.~S.~A.}}
\begin{document}
\begin{abstract} 
Combining (1) the universal 
correlations between $T_{c}$ and $n_{s}/m^{*}$ (superconducting 
carrier density / effective mass) and (2) the pseudo-gap behavior in 
the underdoped region,
we obtain a picture to describe superconductivity in cuprate
systems in evolution from Bose-Einstein to BCS condensation.
Overdoped and Zn-substituted cuprate systems show signatures of
reduced superfluid density in a 
microscopic phase separation.  Scaling of $T_{c}$ to the superfluid 
volume density $n_{s}$ in all these cases indicate importance of 
Bose condensation.
\end{abstract}
\maketitle
The magnetic field penetration depth $\lambda$  
reflects the spectral density
of superfluid as $1/\lambda^{2} \propto n_{s}/m^{*}$ (the superconducting
carrier density / effective mass).  The absolue values of $\lambda$ can 
be determined accurately by Muon Spin Relaxation ($\mu$SR) measurements
where the relaxation rate $\sigma \propto 1/\lambda^{2} \propto n_{s}/m^{*}$
if $H_{c1} \ll H_{ext} \ll H_{c2}$.  Figure 1 shows our $\mu$SR results
in various cuprate and other superconductors in a plot of 
$T_{c}$ versus $\sigma(T\rightarrow 0)$ [1-3].
\begin{figure}
\begin{center}\mbox{\epsfig{file=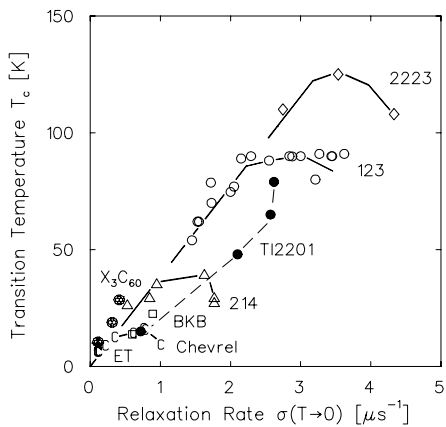,height=5.5cm}}\end{center}\vspace*{-1.cm}
\caption{Muon spin relaxation rate $\sigma(T\rightarrow 0)$ in various
superconductors plotted versus $T_{c}$ [1-3].}
\label{univ}
\end{figure}
This figure demonstrates
that: (A) $T_{c}$ is proportional to $n_{s}/m^{*}$
($m^{*}$ predominantly reflects the in-plane value $m^{*}_{ab}$
in cuprates) 
in the underdoped region 
with the slope common to many 
different cuprates (123, 214, 2223 systems etc.) 
-- universal correlations [1];
(B) $T_{c}$ shows saturation in the ``optimum doping'' region [1];
(C) $n_{s}/m^{*}$ {\it decreases\/} with increasing carrier doping in the 
overdoped region (Tl2201) [3,4]; and (D) not only the cuprates but also 
organic 
BEDT, doped C$_{60}$ and some
other superconductors have ratios between $T_{c}$ and 
$n_{s}/m^{*}$ [2] comparable to those of the cuprates.

Implications of Fig. 1 become clearer if we convert the horizontal axis
into an ``effective'' Fermi energy $\epsilon_{F}$ [2], with $\epsilon_{F}$ in 
2-d systems obtained as $\epsilon_{F} \propto n_{s2d}/m^{*}
\propto \sigma \times c_{int}$ where $n_{s2d}$ stands for the areal
2-d carrier density on each CuO$_{2}$ plane and $c_{int}$ the average
interplanar distance between the CuO$_{2}$ planes.  For 3-d systems,
$\sigma$ can be combined with the Sommerfelt constant $\gamma \propto
n^{1/3}m^{*}$ to obtain $\epsilon_{F} \propto \sigma^{3/4}\gamma^{-1/4}$.
\begin{figure}
\begin{center}\mbox{\epsfig{file=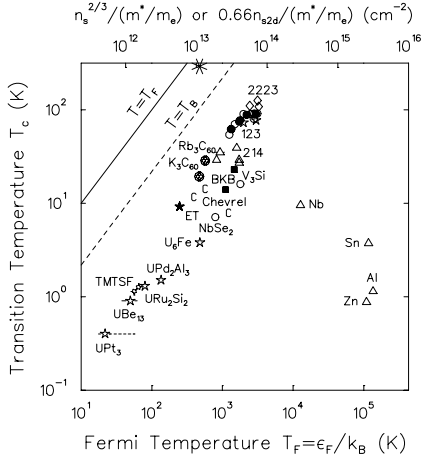,height=6.0cm}}\end{center}\vspace*{-1cm}
\caption{A plot of $T_{c}$ versus the effective Fermi energy $\epsilon_{F}$
calculated from $\sigma$ [2].
$T_{B}$ denotes BE condensation temperature for non-interacting bosons
with density $n_{s}/2$ and mass $2m^{*}$.
The * symbol shows $T_{\theta}^{max}$ of ref. [14]
for K$_{3}$C$_{60}$.}
\label{fermi}
\end{figure}
Figure 2 shows the resulting plot, together with the broken line which
corresponds to the  
Bose-condensation temperture $T_{B}$ for 3-dimensional non-interacting
bosons of boson density $n_{s}/2$ and mass $2m^{*}$.
  
This figure demonstrates that: (E) cuprates, organic BEDT,
and some other ``exotic'' suprconductors have $kT_{c}/\epsilon_{F}$ as high
as 0.01-0.1, much higher than those of conventional BCS superconductors;
(F) the linear relationship $T_{c} \propto n_{s}/m^{*}$ may originate from 
that of Bose-Condensation, as the points lie parallel to the broken line;
(G) $T_{c}$ of cuprates is about 4-5 times smaller than $T_{B}$.
The feature (G) can be attributed partly to the overlap of bosons,
as several pairs exist per $\xi_{ab}^{2}$ ($\xi_{ab}$ is the in-plane
coherence length) on the conducting planes in the cuprates.  
Even for liquid $^{4}$He, finite size
of bosons and their interactions make the lambda transition at 
$T_{\lambda} = 2.2$ K significantly reduced from
$T_{B} \sim 3.2$ K.   
The 2-dimensional character of the cuprates would
further reduce $T_{c}$, as we discuss later.
Figure 2 provides a phenomenological method to classify superconductors in 
evolution from Bose-Einstein (BE) condensation (in real space
with non-retarded strong interaction: close to the broken line) and BCS 
condensation (in momentum space with retarded weak interaction: at large 
$\epsilon_{F}$ and moderate $T_{c}$).

As an interpretation of these results, we presented a picture 
for the high-$T_{c}$ cuprates in terms of  
a crossover from BE to BCS condensation in 1994
[5,6].  This picture stems from the following observations and 
concepts [5,6]:  (A) and 
(F) mentioned above; (H) the loss of mangtic response below the pseudo-gap
temperature $T^{*}$ observed by NMR and neutrons [7] may infer 
a formation of singlet pairs (i.e. pre-formed bosons); 
(I) the insulating behavior observed in c-axis conductivity
below $T^{*}$ [8] 
may be due to reduction of hopping/tunneling probability
between CuO$_{2}$ planes for a 2e particle compared to that 
of a single fermion.
\begin{figure}
\begin{center}\mbox{\epsfig{file=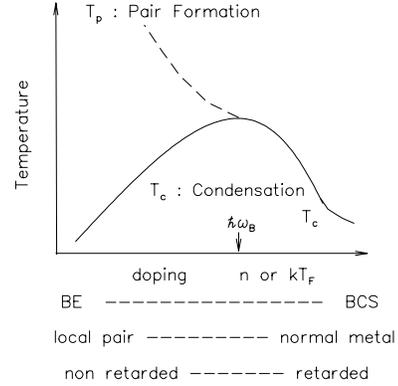,height=5.cm}}\end{center}
\vspace*{-1cm}
\caption{Phase diagram describing BE-BCS crossover with increasing carrier
concentration $n$.  This phase diagram can be mapped to that of the cuprates
by assuming that the pseudo gap temperature $T^{*}$ corresponds to the 
formation of normal-state pairs. [5,6]}
\label{bebcs}
\end{figure}
Then the situation in the cuprates can be mapped to a more general
phase diagram in Fig. 3 [5,6] expected for BE-BCS
crossover with increasing carrier doping.  

When the energy scale of attractive
interaction is higher than $T_{B}$ in the low density region, the singlet
pairs will be first formed at $T^{*}$ and then they undergo BE condensation
at a much lower temperature.  With increasing carrier doping, $T_{c}$ will
increase as $T_{c} \propto T_{B}$ while $T^{*}$ would decrease.  $T_{c}$ and
$T^{*}$ would merge in the ``crossover region'', which corresponds to the 
``optimum $T_{c}$'' region of the cuprates.
The higher density side is characterized by the simultaneous
pair-formation and condensation at $T_{c}$, i.e., 
the most fundamental feature of
BCS condensation.  We further conjectured that (J) the crossover
will occur where $\epsilon_{F}$ becomes comparable to the energy scale
$\hbar\omega_{B}$ of the exchange bosons which mediate the pairing interaction
[5,6], thus characterizing the underdoped cuprates with retarded coupling.
Assuming (J), we can estimate $\hbar\omega_{B}/k_{B}$ to be comparable to 
$\epsilon_{F}/k_{B} \sim 2,000$ K of the ``optimally doped'' cuprates.
We recently pointed out [9] that: (K)
this energy scale matches well with that of the antiferromagnetic 
exchange interaction
$J$, providing a support for pictures that spin fluctuations indeed 
mediate the pairing interaction; and (L) the mid infrared reflection 
(MIR) in optical conductivity of the cuprates, 
appearing with energy scales comparable to 
$\hbar\omega_{B}$, could originate from the pairing interaction,
in [6] contrast to the Drude part representing the translational motion 
of carriers (with an energy scale $\epsilon_{F}$).  

This picture is re-inforced by recent observation of photo-emission
studies [10] that the superconducting gap in the ``optimal'' $T_{c}$
region evolves smoothly to the pseudo gap in the underdoped region
with the same geometrical symmetry.  
General concepts of negative-U Hubbard model and 
BE-BCS crossover have been considered by number of authors
[11,12].  In 1990, Doniach and Inui pointed out that the pseudo gap
behavior may be interpreted in terms of phase fluctuations of the 
superconducting order parmeter [13].    
In 1995, Emery and Kivelson [14] discussed that the universal correlations
between $T_{c}$ and $n_{s}/m^{*}$ (A) is expected when $T_{c}$ is determined
by the phase fluctuations.
Although they argued this as ``re-interpretation'' of (A), 
the present author
believes that their phase-fluctuation picture is essentially identical to 
our BE-BCS crossover picture.  

Emery and Kivelson calculated the kinetic energy for phase fluctuations,
and estimated the maximum condensation temperature 
$T_{\theta}^{max} \propto n_{s}/m^{*} \times y$, where $y$ is a
parameter representing distance.  For 2-d systems they substituted 
the interplanar distance $c_{int}$, which lead to 
$T_{\theta}^{max} \propto T_{B}$. Their choice of the coherence length
$\xi$ as $y$ for 3-d systems led to unrealistically high value of 
$T_{\theta}^{max} \gg T_{B}$, as shown in Fig. 2 for the example of 
K$_{3}$C$_{60}$.  (Note that for $\xi \rightarrow \infty$,
$T_{\theta}^{max} \rightarrow \infty$.) 
The ultimate upperlimit for condensation temperature should be given by
$T_{B}$, since any interaction, dimensionality, and other factors would
reduce actual $T_{c}$ from $T_{B}$.  These considerations
indicate 
that $y$=$\xi$ for 3-d systems in ref. [14],
which led to an apparent difference between 2-d and 3-d systems in their 
Table 1, is a bad choice.  
Instead, if one substitutes the interparticle 
distance for $y$ for 3-d systems, Table 1 of ref. [14] reduces to 
Fig. 3 of ref. [2].

The role of the 2-dimensional nature of cuprates is an 
interesting issue.  Figure 1 does not give much information on this, 
as the horizontal axis represents the 3-d carrier
density $n_{s}$ as well as the 2-d density $n_{s2d} = n_{s} \times c_{int}$
for the 123, 214 and 2223 
systems, all having the interlayer distance $c_{int} \sim 6$\AA. 
Recently, we have performed $\mu$SR measurements in Hg1201 [15]
systems, 
with $c_{int}$ = 9.5 \AA. Our previous data of nearly-optimally doped Tl2201 
system [3] represents the case with $c_{int}$ = 11.5 \AA.  
All these results and those of underdoped
214, 123 and 2223 systems lie approximately on 
the same straight line when we plot
$T_{c}$ versus $\sigma \propto n_{s}/m^{*}$.  However, 
when plotted versus $\sigma \times c_{int} \propto n_{s2d}/m^{*}$ in 
Fig. 4, we find that: (M) systems with smaller $c_{int}$ have 
higher $T_{c}$ for a given
$n_{s2d}/m^{*}$, following a relation $T_{c} \propto 1/c_{int}$.
\begin{figure}
\begin{center}\mbox{\epsfig{file=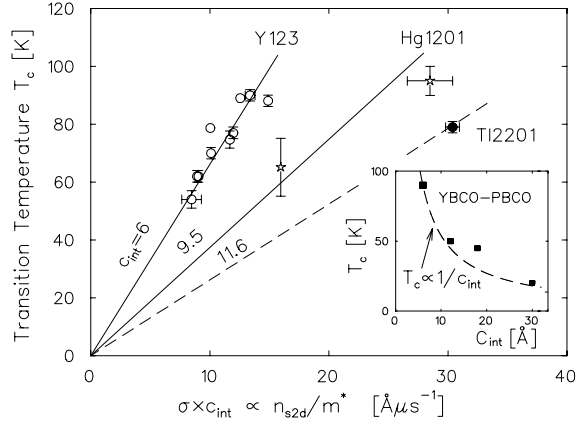,height=5.5cm}}\end{center}\vspace*{-1.cm}
\caption{Plot of $T_{c}$ versus $\sigma \times c_{int}
\propto n_{s2d}/m^{*}$ for the underdoped region of Y123 [1], Hg1201 [15]
and neally-optimally doped Tl2201 [3] 
systems.  Inset: $T_{c}$ vs. $c_{int}$ in YBCO-PBCO [16].} 
\label{univ2d}
\end{figure}
This result is consistent with 
the reduction of $T_{c}$ for increasing thickness of 
PrBa$_{2}$Cu$_{3}$O$_{7}$ (PBCO) layer   
in thin films of  
YBa$_{2}$Cu$_{3}$O$_{7}$ (YBCO) having 
a single unit-cell stacked in the c-axis direction
with non-superconducting PBCO [16] (see Fig. 4).

This feature $T_{c} \propto n_{s2d}/c_{int}$ (=$n_{s}$) 
indicates: 
(N) a Kosterlitz-Thouless transition in its simplest
form, where $n_{s2d}/m^{*}$ is the sole factor for $T_{c}$, in not sufficient
for predicting $T_{c}$; 
and 
(O) BE condensation in quasi 2-d systems gives a better explanation 
since $T_{c}$ dependes 
logarithmically as 
$T_{c} \propto n_{s2d}/m^{*} \times [1/\ln(1/I_{c})]$ 
on the interlayer tunnelling probability $I_{c}$ [17]
while $I_{c}$ should depend exponentially on $c_{int}$
as $I_{c} \propto \exp(-c_{int})$, leading to $T_{c} \propto 
1/c_{int} \times n_{s2d}/m^{*}$.  

The decrease of $n_{s}/m^{*}$ with increasing doping 
in the overdoped Tl2201 systems (feature (C) in Fig. 1) 
can {\it not\/} be expected in the 
BCS model with retarded interaction, where the 
nomal state carrier density $n$ should be equal to $n_{s}$.
In ref. [3], we proposed (P) a spontaneous microscopic phase separation
between superconducting and normal regions ($n_{s} < n$)
as a possible explanation.  
Recently, we found a similar situation in Zn-substituted 
214 and 123 systems [18], where 
the reduction of $n_{s}/m^{*}$
with increasing Zn doping suggests 
that: (Q) carriers within the region $\pi\xi_{ab}^{2}$ on the CuO$_{2}$
planes around each Zn impurity may be excluded from the superfluid --
like ``swiss cheese''.  
Specific heat in overdoped Tl2201 and Zn-doped 123/214 systems both
exhibit increasing T-linear term at $T \rightarrow 0$ 
with increasing carrier/Zn doping [19],
further supporting our models (P) and (Q) with 
phase separation.

The short coherence
length $\xi$ in cuprates will help reduce the energy cost for 
creating phase boundaries, facilitating phase separation.
Indeed the swiss cheese model for Zn-doped systems is reminiscent of
the behavior of superfluid $^{4}$He in porous media [20].
For overdoped cuprates, we propose the following
scenario: (R)  
if an abrupt decrease of effective attractive interaction
occurs in the cuprates as the 
doping exceeds the ``optimal'' concentration, then 
energy gain for maintaining superconductivity 
might win over
energy cost for phase separation:   
the ``cheese'' part 
composed of regions with local $n_{s}$ 
near the ``optimum concentration'',
while its volume fraction decreasing as further doping.

In underdoped (M),
overdoped (P) and Zn-doped (Q) cuprates,
$T_{c}$ is proportional to   
the (spatially averaged) 3-dimensional volume density $n_{s}$
of superfluid.  This demonstrates fundamental 
importance of BE condensation. 

This study is supported by NSF (DMR-95-10453, 10454)
and NEDO (Japan).  The author is indebted to Oleg Tchernyshyov for 
discussions leading to the interpretations (N) and (O).

\end{document}